\documentclass[epj]{svjour}
\usepackage{latexsym}
\usepackage{graphics}
\usepackage{float}

\begin{document}
\title{Time travel, Clock Puzzles
and Their Experimental Tests}

\author{Ignazio Ciufolini\inst{}
}
%
%
\institute{Dip. Ingegneria dell'Innovazione, Universit\`a del Salento, Lecce, and Centro Fermi, Rome, Italy \\
  \email{ignazio.ciufolini@unisalento.it}}
%
%
\abstract{
Is time travel possible? What is Einstein's theory of relativity mathematically predicting in that regard?
Is time travel related to the so-called clock `paradoxes' of relativity and if so how? Is there any accurate experimental
 evidence of the phenomena regarding the different flow of time predicted by General Relativity and is there any possible
 application of the temporal phenomena predicted by relativity to our everyday life? Which temporal phenomena are predicted
 in the vicinities of a rotating body and of a mass-energy current, and do we have any experimental test of the occurrence
 of these phenomena near a rotating body? In this paper, we address and answer some of these questions.}

%
%
\authorrunning{ Ciufolini}
\maketitle

\section{Introduction to time travel and clock puzzles and paradoxes in relativity}
\label{intro}

Is time travel possible? In order to attempt to answer this question we should first divide it into two parts: ``Can we travel into the future?" and ``Can we travel into the past?"
If by travel into the future, we mean the situation whereby, under certain conditions, a clock or an observer, may be in the future of other clocks and observers, on Earth or elsewhere,
 for example, one of us could find himself, after just one year of his biological time, also marked by his wristwatch, in the year 2100 of the rest of humankind, then the answer is yes.
Indeed, although not perceiving it, many of us, under certain conditions, have traveled in time without knowing it, for example after a flight. The fact that the time difference between our clock
 and other clocks on Earth was only the order of a few nanoseconds is `only' a quantitative difference, compared with the journey in the year 2100, and not qualitative. This is difference
which we do not perceive but an atomic clock can perceive and can accurately measure.

Now we come to the second question ``Is time travel to the past possible"? General Relativity \cite{mtw,ciuw} suggests that this
is mathematically possible, indeed it actually predicts the existence of closed timelike lines \cite{time0}, for example in the first G\"odel cosmological models \cite{chand} and in the Kerr metric \cite{kerr}. If we could follow
one of these lines, we could go back in time. Nevertheless, apart from the mathematical existence of closed timelike lines in General Relativity, we do not know with certainty the answer in the
`real' universe. This possibility is discussed by physicists and cosmologists like Steven Hawking, Igor Novikov, Kip Thorne, et al. \cite{time0,time1,time2,time3,time4,time5,time6,time7,time8,time9,time10,time11}, and involves numerous physical and logical problems,
 which are briefly described in section 4.

Time travel to the future is related to the so-called `paradoxes' of Special and General Relativity, such as the well-known `twin paradox'. It would be better to call them `time puzzles', however,
 since they do not imply any logical paradox but only situations that are difficult to imagine. Some of these `time puzzles' have been confirmed by a number of very accurate experiments.
 Some `time puzzles' of General Relativity, such as the time dilation of a clock near a mass, involve striking effects in the strong gravitational field of a black hole. Indeed, although
the name `black hole' was proposed in the USA by John Archibald Wheeler, his colleagues in Russia called such bodies `frozen stars' because a clock near the event horizon of a black
 hole as seen by a distant observer would appear as almost still or frozen. Nevertheless, this phenomenon has been confirmed in the weak gravitational field of the solar system by
a number of very accurate experiments which are described in the next section.

In the first part of this paper we describe the main experimental tests of some of the `time puzzles' of General Relativity. Some of these tests relate to our everyday
life and, probably without knowing it, we test these clock effects whenever we use a GNSS (Global Navigation Satellite System) navigator.

In the second part of this
work we describe some clock phenomena of General Relativity that involve a current of mass-energy or a rotating mass. These clock effects are the result of the
 General Relativistic phenomenon of `frame-dragging' or `dragging of inertial frames' \cite{ciu07}, also called `gravitomagnetism' in a weak gravitational field \cite{tho}. Finally,
 we describe the main experimental tests of frame-dragging obtained with the two LAGEOS satellites \cite{ciu04,ciu07,rie09,ciu10,ciu11} and with Gravity Probe B \cite{GPB}, and we describe the space experiment with the LARES satellite \cite{ciu10b,ciu12,ciupao}, successfully launched on 13 February 2012, for very accurate tests of frame-dragging.

\section{Time travel to the future via clock time dilation by a
mass and its experimental tests}

The time dilation of a clock in a gravitational field, or gravitational red-shift, may be considered as a consequence of the equivalence principle. It may in fact be derived in a weak field
 from the medium strong equivalence principle (valid in any metric gravity theory), conservation of energy and basic classical and quantum mechanics \cite{ein}.
If we consider, for simplicity, a static spacetime, we can then find a coordinate system where both the metric is time independent: $g_{\alpha \beta , 0} = 0$ and its non-diagonal
 components are zero: $g_{i0} = 0$, and therefore the coordinate time $t$ required for an electromagnetic signal to go from a coordinate point $A$ to any other coordinate point $B$
 is the same as the coordinate time $t$ for the signal to return from $B$ to $A$. One can then consistently define simultaneity \cite{lanlif} between any two events
 using light signals between them.
By the equivalence principle, the interval of proper time $\tau$ measured by a clock at rest in a freely falling frame, at an arbitrary point in the gravity field,
 is related to the coordinate time between two events, at $x^{\alpha} = constant$, by:
\begin{displaymath}
d \tau^2 = g_{\alpha \beta} \, dx^\alpha \, dx^\beta = - {g_{0 0}}^{1/2} \, dt
\end{displaymath}
Since an interval of coordinate time represents the same interval of coordinate time between simultaneous events all over the manifold, a phenomenon that takes a proper time: $\Delta \tau _A \, \simeq \, ( \, - g_{0 \, 0} \, (A) \, ) \,^{1/2} \, \Delta  t$, in $A$, as seen by one observer in $B$ takes the proper time $\Delta \tau_B \,  \simeq \, (\, - g_{0 \, 0} \, (B) \, ) \, ^{1/2} \, \Delta  t$, that is at the lowest order:
\begin{equation}
\Delta \tau _B = \Bigl ( \, {g_{0 \, 0} \, (B) \over g_{0 \, 0} \, (A)} \, \Bigr )^{1/2} \, \Delta \tau _A
\simeq \Delta \tau_A \, ( 1 + U_A - U_B ) \equiv \Delta \tau _A ~ ( 1 - \Delta U )
\end{equation}
where $U$ is the classical gravitational potential, therefore if $r_A > r_B ,~ \Delta \, U \equiv U_B - U_A \, > 0$, and the duration of a phenomenon measured by the proper time of a clock in B: $\Delta \tau_B$ , appears  longer when measured by the proper time of a clock in A: $\Delta \tau_A$, that is, a clock near a mass appears to go slower as observed by clocks far from the mass. This is the time dilation of clocks in a gravity field equivalent to the gravitational red-shift, that is the frequency red-shift of an electromagnetic wave that propagates from a point near a mass to a more distant point.

In general, the gravitational time dilation of clocks, or gravitational red-shift, in any frame of any spacetime, for any two observers, emitter and detector, moving with arbitrary four-velocities, is given by:
\begin{equation}
{ ( {d \tau }) \over ({d \tau^\ast })} ~ = ~  {g_{\alpha \beta} \, u^\alpha \, d x^\beta \over g_{\alpha \beta}^\ast \, u^{\ast \alpha} \, {d x^{\ast \beta}}}
\end{equation}
where the observer with four-velocity $u^\alpha$, with metric $g_{\alpha \beta}$, measures the proper time $d \tau$ corresponding to the events: $x^\alpha$ (beginning of emission), and $x^\alpha + d \, x^\alpha$ (end of emission) and where the $^\ast$ refers to the four velocity, metric and coordinates of beginning and end of detection at the detector.

The gravitational time dilation of clocks has been tested by numerous accurate experiments \cite{will,ciuw}. Among the red-shift experiments measuring the frequency shift of electromagnetic waves propagating in a gravity field, we recall the `classical' Pound--Rebka--Snider experiment that measured the frequency shift of $\gamma$ rays rising a 22.6 meter tower at Harvard with an accuracy of about 1 $\%$. Other tests measured the gravitational red-shift from the Sun, from white dwarfs, from Sirius and other stars. The gravitational red-shift owed to Saturn was measured by the Voyager 1 spacecraft during its encounter with Saturn and that owed to the Sun by the Galileo spacecraft and using the millisecond pulsar PSR 1937+21. Gravitational red-shift is one of the relativistic parameters measured with high accuracy using the 1974 binary pulsar B1913+16.

Other `classical' experiment include the Hafele--Keating \cite{hafkea1,hafkea2} test in 1972, using cesium--beam clocks on jets that confirmed the predictions of Special and General Relativity with an accuracy of about 20 $\%$ and the Alley test in 1979, using rubidium clocks on jets, with accuracy of about 20 $\%$.
The most accurate {\it direct} time dilation measurement so far, however, is the Vessot and Levine \cite{vess1,vess2} (1979--80) NASA clock experiment, called GP-A (Gravity Probe A). Two hydrogen--maser clocks, one on a rocket at about 10,000 km altitude and one on the ground, were compared. The accuracy reached, in agreement with the theoretical prediction of General Relativity, was 2 $\times \, 10^{-4}$. The general relativistic clock dilation predicts that for every second of the clock on the spacecraft, the clock on Earth is retarded by about 4 $\times \, 10^{-10}$ seconds.

Future tests of gravitational time dilation will include the space mission ACES with atomic clocks on the International Space Station (ISS) \cite{aces}, that should reach an accuracy of about 2 $\times \, 10^{-6}$ in testing gravitational time dilation, and the ESA space experiment SOC (Space Optical Clock) using optical clocks on ISS that should reach an accuracy of about 2 $\times \, 10^{-7}$.
 A proposed interplanetary experiment would compare high precision clocks on Earth with similar clocks orbiting near the Sun, at a few solar radii; its accuracy should be of the order of 10$^{-9}$ and it should then be able to test second order time dilation effects in the Sun gravitational potential.

\section{Clock effects and everyday life}

In the previous section we have seen that the difference between a clock on the ground and a clock in space, a few hours after their synchronization, may be a few microseconds. The time difference between a clock on the ground and a clock on an airplane, after a few hours, may be a few nanoseconds. Thus, most of us have experienced `time travel' for such small periods of time but of course we could not `feel' such a small time difference; nevertheless an atomic clock can `feel' it and, indeed, atomic clocks have measured the time dilation of clocks with an accuracy of about two parts in ten thousand and in the near future they will test it with accuracy that may reach about two parts in ten millions (see the previous section).

These relativistic phenomena regarding the `flow of time' and clocks are extremely interesting from a theoretical and astrophysical point of view but, since they involve very small time differences, they appear to be very far from our everyday life. Therefore can these temporal phenomena ever have any practical effect on our everyday lives? Yes, they can and amazingly they already have important daily applications.

Today, by using a satellite navigator, we can find our position with an error of just a few tens of meters and for that we have also to thank the clock corrections predicted by General and Special Relativity. If the relativistic corrections of the different flow of time at different distances from a mass and those owed to the speed of a clock were not included in the calculations of the position of the satellites of the global navigation systems, then there would be errors in our positioning of the order of kilometers \cite{ash}. The satellites of the USA Global Positioning System (GPS) are orbiting at an altitude of about 20,000 km and their atomic clocks, because of the combined clock effects of General and Special Relativity, are faster than those on the ground by approximately 39 microseconds per day. If we multiply this interval of time for the speed of light, and if we do not take into account the relativistic effects on the flow of time, there would be an error in our positioning amounting after a few hours, to kilometers! It is amazing that in our everyday life, without knowing it, we put to the test the time dilation of clocks predicted by relativity whenever we use a satellite navigator!

The current global navigation satellite systems (GNSS) are the USA GPS (Global Positioning System) and the Russian GLONASS. China is deploying the Compass global satellite navigation system and the European Union the GALILEO global navigation system which is planned to be fully operational in 2020. Four satellites of the GALILEO constellation plus two test satellites are already in orbit (see Fig. 1). The fully operational GALILEO system will consist of a total of 30 satellites arranged in such a way that at any time at least four satellites will be visible anywhere on Earth. The last two satellites of the GALILEO constellation were successfully launched in 2012 from Baikonur with a Soyuz rocket.

\begin{figure}
\centering \resizebox{0.55\textwidth}{!}{
  \includegraphics{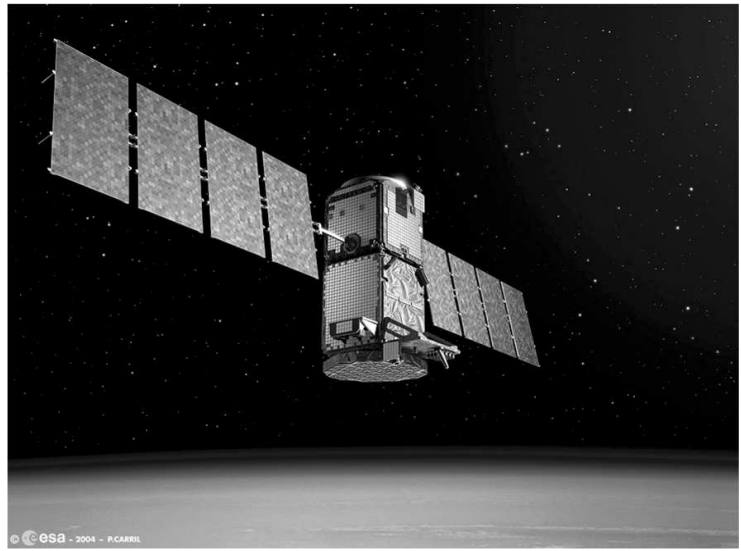}
} \caption{GIOVE B, a test satellite of the European GALILEO constellation with the most accurate clock (in 2008) ever launched into space}
\end{figure}

\section{Time travel to the past?}

Is time travel to the past possible? The Einstein's field equations of General Relativity predict the mathematical existence of closed timelike lines, such as those in the Kerr metric and in the first G\"odel cosmological models and the mathematical existence of wormholes, such as the Einstein-Rosen bridge \cite{mtw}, connecting distant regions of the universe. Closed timelike lines are `time machines'. An observer who was able to follow a closed timelike line could return back to the same starting spacetime event, i.e., could go back in time. Apart from the existence of closed timelike lines as mathematical solutions of the Einstein's field equations, however, we do not know if time travel to the past would be possible in the `real' universe.

A wormhole can be turned into a `time machine' if we place for example one of its two mouths in a gravitational field, e.g., that of a neutron star star \cite{time7}. Then, by traveling through it from that mouth closer to the neutron star one could go back in time. Nevertheless, to be able to keep a wormhole open, i.e., to gravitationally push the walls of the wormhole apart, one would need to use some `exotic matter' with negative average energy density \cite{time7}. Furthermore, the possibility of going back in time seems to imply the so-called grandfather or `grandmother paradox'.
An observer could go back in time using a wormhole and kill his grandmother before his mother was born, therefore preventing his own birth and the killing of the grandmother. That paradox was translated in one, less bloody, example involving a billiard ball entering a wormhole, going back in time and hitting itself in the past in such a way that its trajectory was no longer able to enter the mouth of the wormhole resulting in a self-consistency paradox.
To avoid the `grandfather paradox', Thorne, Novikov \cite{time3,time4,time5,time6,time7} et al. have shown that in the case of the billiard ball, for example, there are always mathematical solutions free from self-inconsistent causality violations like the `grandfather paradox'. For example in the case of the billiard ball there are self-consistent solutions with the ball going through the wormhole even after being hit by itself coming back from the future. To prevent the occurrence of closed timelike curves and ``to make the world safe for historians", Hawking \cite{time8} formulated the `Chronology Protection Conjecture'. Detailed discussions about the possibility of time travel and of some advanced civilization being able to build a `time machine' are presented elsewhere in this book and in \cite{time0,time1,time2,time3,time4,time5,time6,time7,time8,time9,time10,time11}.

\section{Clock puzzles owed to frame-dragging around a rotating body}

There is another type of `clock puzzle' that occurs near a rotating body or a current of mass-energy, that resulting from a relativistic phenomenon, described in the next section, called `frame-dragging' or `dragging of inertial frames'.
Frame-dragging has an intriguing influence on the flow of time around a spinning body (Fig. 2). Indeed, synchronization of clocks all around a closed path near a spinning body is not possible \cite{lanlif,zeln} in any rigid frame not rotating relative to the `fixed stars', because light co-rotating around a spinning body would take less time to return to a starting point (fixed relative to the `distant stars') than would light rotating in the opposite direction \cite{lanlif,zeln,ciur1,ciur2,ciurk,tar1,tar2,tar3}.

\begin{figure}
\centering \resizebox{0.45\textwidth}{!}{
  \includegraphics{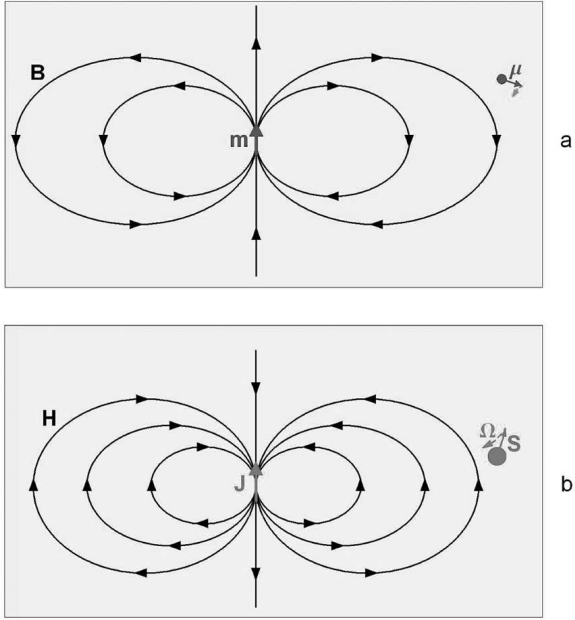}
} \caption{Two pulses of radiation counter-propagating in a circuit where $g_{0i} \ne 0$. $~ \oint ~  {g_{0i} \over g_{00}} ~\scriptstyle d \, \scriptstyle x^i \sim {J \over r}$ for a gravitomagnetic field  $g_{0i} \sim {J \over r^2}$, and $\oint ~ {g_{0i} \over g_{00}} ~ \scriptstyle d \, \scriptstyle x^i~ \sim ~ \dot \Omega \, r^2$ for a circuit rotating with angular velocity $\dot \Omega$}
\end{figure}

In every stationary spacetime we can find a coordinate system such that the metric is time independent:  $g_{\alpha \beta , 0} = 0$, but only if the spacetime is static would the non-diagonal components, $g_{0i}$, of the metric be zero in that coordinate system. Then, two events with coordinates $x^\alpha$ and $x^\alpha  + d \, x^\alpha$, are simultaneous for an observer at rest in that coordinate system if: $d \, x^0 = \, - {g_{0i} ~ d \, x^i   \over  g_{00}}$.
This condition of simultaneity between two nearby events corresponds to sending a light pulse from $x^i$ to $x^i + d \, x^i$, then reflecting it back to $x^i$, and to defining at $x^i$ the event simultaneous to the event of reflection at $x^i + d \, x^i$, as that event with coordinate time, at $x^i$, halfway between the pulse departure time and its arrival time back at $x^i$. Using this condition, one can synchronize clocks at different locations, but if $g_{0i}$ is different from zero, that condition of simultaneity along a path involves $d \, x^i$, i.e., the synchronization between two clocks at different locations depends on the path followed to synchronize the clocks. Therefore, if $g_{0i}$ is different from zero, by returning to the initial point along a closed circuit one gets in general a non-zero result for simultaneity, i.e., the integral $\oint ~ {g_{0i} ~  d \, x^i   \over  g_{00}}$ is in general different from zero, and consequently, one cannot consistently define simultaneity along a closed path.

The physical interpretation of this effect is that, since to define the simultaneity between two events one uses light pulses, a different result for simultaneity, depending from the spatial path chosen, means an interval of time for a pulse to go to a point different from the interval of time for the pulse to return back to the initial point along the same spatial path \cite{lanlif,zeln,ciur1,ciur2,ciurk}. This effect depends on the path chosen and is owed to the non-zero $g_{0i}$ components. In general, for a metric tensor with non-diagonal components different from zero, such as the metric generated by a rotating body, the interval of time for light to travel in a circuit is different if the light propagates in the circuit in one sense or the other. For example, around a central body with angular momentum $J$, the spacetime is described, in the weak field limit, by the Lense-Thirring metric (i.e., the Kerr metric for weak gravitational field and slow motion of the source) and the difference is proportional to $\oint ~ {g_{0i} \over g_{00}} ~\scriptstyle d \, \scriptstyle x^i ~ \sim ~ J / r$ (see Fig. 2).

In Fig. 3 is described a clock `puzzle' owed to the spin of a central body. For this effect to occur, the clocks, or twins, would not need to move close to the speed of light (as in the case of the well-known `twin-paradox' of special relativity). For example, if two such twins meet again, having flown arbitrarily slowly around the whole Earth in opposite directions on the equatorial plane and exactly at the same altitude, the difference in their ages owed to the Earth's spin would be approximately 10$^{-16}$ s (for an altitude of about 6,000 km), which would in principle be detectable if not for the other, much larger, relativistic clock effects. These clock effects are striking around a rotating black hole, however.

\begin{figure}
\centering \resizebox{0.45\textwidth}{!}{
  \includegraphics{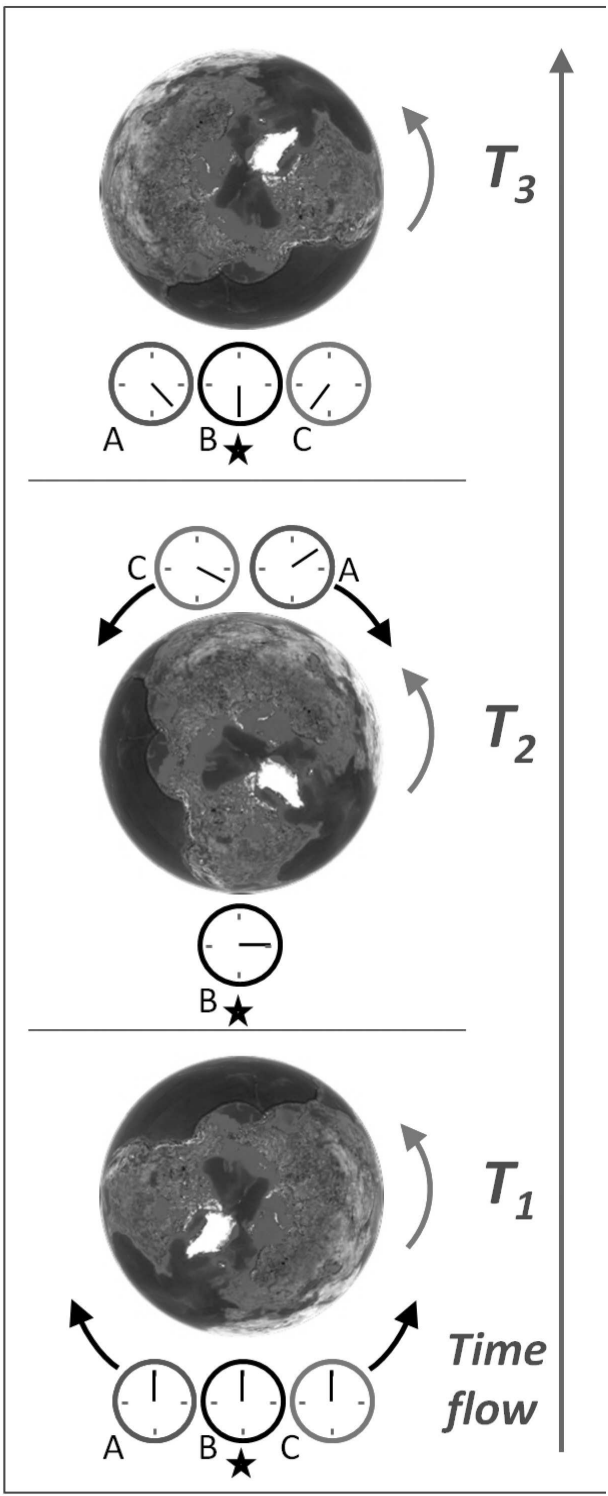}
} \caption{The lack of the possibility of consistently defining simultaneity around a rotating body implies the following clock puzzle. If two  clocks, or twins A and C, go all around a spinning body, very slowly, and a third one B awaits them at the starting-point fixed relative to the `distant stars' (a `fixed star' is shown in black, and T1, T2 and T3 are three consecutive instants of time), then when they meet again the twin A that was traveling in the direction opposite to the rotation of the central body, would be younger than the twin B waiting at the starting-point. On the other hand, twin C, traveling in the same direction of rotation of the body, would be older with respect to the standing twin B and to the twin A rotating in the opposite direction \cite{lanlif,zeln,ciur1,ciur2,ciurk} (Earth's image by NASA and Google Earth)}
\end{figure}

In the next section, we shall see that frame-dragging produces relevant effects not only on clocks and light but also on matter, test-particles and test-gyroscopes around a rotating body or near a current of mass-energy.

\section{Frame Dragging}

In Einstein's gravitational theory the local inertial frames have a key role \cite{mtw,ciuw,wei}. The strong equivalence principle, at the foundations of General Relativity, states that the gravitational field is locally `unobservable' in the freely falling frames and thus, in these local inertial frames, {\it all} the laws of physics are the laws of Special Relativity. The medium strong equivalence principle, valid for any metric theory of gravitation \cite{wei,ciuw,will}, states that in the local inertial frames, all the non-gravitational laws of physics are the laws of Special Relativity \cite{wei}. However, the local inertial frames are determined, influenced and dragged by the distribution and flow of mass-energy in the Universe. The axes of these non-rotating, local, inertial frames are determined by free-falling torque-free test-gyroscopes, i.e., sufficiently small and accurate spinning top. Therefore, these gyroscopes are dragged by the motion and rotation of nearby matter \cite{mtw,ciuw,wei}, i.e., their orientation changes with respect to the `distant stars': this is the `dragging of inertial frames' or `frame-dragging', as Einstein named it in a letter to Ernst Mach \cite{mach}. If we rotated with respect to dragged gyroscopes, we would then feel centrifugal forces, even though we may not rotate at all with respect to the `distant stars', contrary to our everyday intuition.

Frame dragging phenomena, which are due to mass currents and mass rotation, have been called gravitomagnetism \cite{tho,ciuw} because of a formal analogy of electrodynamics with General Relativity (in the weak-field and slow-motion approximation). Whereas an electric charge generates an electric field and a current of electric charge generates a magnetic field, in Newtonian gravitational theory the mass of a body generates a gravitational field but a current of mass, for example the rotation of a body, would not generate any additional gravitational field. On the other hand, Einstein's gravitational theory predicts that a current of mass would generate a gravitomagnetic field that would exert a force on surrounding bodies and would change the spacetime structure by generating additional curvature \cite{kerr,ciuw,ciu1}. The gravitomagnetic field generates frame-dragging of a gyroscope, in a similar way to the magnetic field producing the change of the orientation of a magnetic needle (magnetic dipole). Indeed, in the general theory of relativity, a current of mass in a loop (that is, a gyroscope) has a behavior formally similar to that of a magnetic dipole in electrodynamics, which is made of an electric current in a loop.

In the previous section, we have seen that frame-dragging around a spinning body has an intriguing influence on the flow of time. Synchronization of clocks all around a closed path near a spinning body is in general not possible and light co-rotating around a spinning body would take less time to return to a `fixed' starting point than would light rotating in the opposite direction.

However, frame-dragging affects not only clocks and electromagnetic waves but also gyroscopes \cite{pug,sch} (e.g., the gyroscopes of GP-B space experiment) and orbiting particles \cite{len} (e.g., the LAGEOS and LARES satellites), for example matter orbiting and falling on a spinning body. Indeed, an explanation of the constant orientation of the spectacular jets from active galactic nuclei and quasars, emitted in the same direction during a time that may reach millions of years, is based on frame-dragging of the accretion disk due to a super-massive spinning black hole \cite{bard,tho} acting as a gyroscope.

\section{Experimental Tests of Frame-Dragging}

Since 1896 researchers, influenced by the ideas of Ernst Mach, tried to measure the frame-dragging effects generated by the rotation of the Earth on torsion balances \cite{frie} and gyroscopes \cite{fop}. In 1916, on the basis of General Relativity, de Sitter derived the Mercury perihelion precession due to the Sun angular momentum and, in 1918, Lense and Thirring \cite{len} gave a general weak-field description of the frame-dragging effect on the orbit of a test-particle around a spinning body, today known as Lense-Thirring effect.

The precession ${\dot {\bf \Omega}}_{Spin}$ of the spin axis of a test-gyroscope by the angular momentum $\bf J$ of the central body is: ${\dot {\bf \Omega}}_{Spin} = {{3 G (({\bf J} \cdot \hat{r}) \hat{r} - {\bf J})} \over {c^2 {r^3}}}$, where $\hat{r}$ is the position unit-vector of the test-gyroscope and r is its radial distance from the central body.

In 1959 and 1960, an experiment to test the general relativistic drag of a gyroscope was suggested \cite{pug,sch}. On 20 April 2004, after more than 40 years of preparation, the Gravity Probe B spacecraft was finally launched in a polar orbit at an altitude of about 642 km. The Gravity Probe B mission \cite{GPB} (see http://einstein.stanford.edu/) consisted of an Earth satellite carrying four gyroscopes and one telescope, pointing at the guide star IM Pegasi (HR8703), and was designed to measure the drifts predicted by General Relativity (frame-dragging and geodetic precession) of the four test-gyroscopes with respect to the distant `fixed' stars. General Relativity predicts that the average frame-dragging precession of the four Gravity Probe B gyroscopes by the Earth's spin will be about 39 milliarcseconds per year, that is, 0.000011 degrees per year, about an axis contained in Gravity Probe B's polar orbital plane.

On 14 April 2007, after about 18 months of data analysis, the first Gravity Probe B results were presented: the Gravity Probe B experiment was affected by unexpected large drifts of the gyroscopes' spin axes produced by unexpected classical torques on the gyroscopes. The Gravity Probe B team explained \cite{buch} (see also \cite{baroco}) the large drifts of the gyroscopes as being due to electrostatic patches on the surface of rotors and housings, and estimated the unmodeled systematic errors to be of the order of 100 milliarcseconds per year, corresponding to an uncertainty of more than 250\% of the frame-dragging effect by the Earth spin. However, in 2011 the Gravity Probe B team claimed that by some modeling of the systematic errors they were able to reduce the uncertainty in the measurement of frame-dragging to 19 \% \cite{GPB}.

Similarly to a small gyroscope, the orbital plane of a planet, moon or satellite is a huge gyroscope that feels general relativistic effects. Indeed, frame-dragging produces a change of the orbital angular momentum vector of a test-particle, known as Lense-Thirring effect, that is, the precession of the nodes of a satellite, i.e., the rate of change of its nodal longitude: ${\dot {\bf \Omega}}_{Lense-Thirring} = {{2 G \bf {J}} \over {c^2 a^3 (1-e^2)^{3/2}}}$, where $\bf \Omega$ is the longitude of the nodal line of the satellite (the intersection of the satellite orbital plane with the equatorial plane of the central body), $\bf J$ is the angular momentum of the central body, $a$ the semi-major axis of the orbiting test-particle, $e$ its orbital eccentricity, $G$ the gravitational constant and $c$ the speed of light. A similar formula also holds for the rate of change of the longitude of the pericentre of a test--particle, that is, of the so-called Runge-Lenz vector \cite{len,ciuw}.

However frame-dragging is extremely small for Solar System objects, so to measure its effect on the orbit of a satellite we need to measure the position of the satellite to extremely high accuracy. Laser-ranging is the most accurate technique for measuring distances to the Moon and to artificial satellites such as LAGEOS (LAser GEOdynamics Satellite) \cite{lag}. Short-duration laser pulses are emitted from lasers on Earth and then reflected back to the emitting laser-ranging stations by retro-reflectors on the Moon or on artificial satellites. By measuring the total round-trip travel time we are today able to determine the instantaneous distance of a retro-reflector on the LAGEOS satellites with a precision of a few millimeters \cite{noon} and their nodal longitude with an uncertainty of a fraction of a milliarcsec per year \cite{nasasi,riedis,pet}.

LAGEOS was launched by NASA in 1976 and LAGEOS 2 was launched by the Italian Space Agency and NASA in 1992, at altitudes of approximately 5,900 km and 5,800 km respectively. The LAGEOS satellites' orbits can be predicted, over a 15-day period, with an uncertainty of just a few centimeters \cite{nasasi,riedis,pet}. The Lense-Thirring drag of the orbital planes of LAGEOS and LAGEOS 2 is \cite{ciu86,ciu89} approximately 31 milliarcseconds per year, corresponding at the LAGEOS altitude to approximately 1.9 m per year. Since using laser-ranging we can determine their orbits with an accuracy of a few centimeters, the Lense-Thirring effect can be measured very accurately on the LAGEOS satellites' orbits if all their orbital perturbations can be modeled well enough \cite{ciu86,ciu89,nasasi}. On the other hand, the LAGEOS satellites are very heavy spherical satellites with small cross-sectional areas, so atmospheric particles and photons can only slightly perturb their orbits and especially they can hardly change the orientation of their orbital planes \cite{ciu89,nasasi,rub,luc}.

By far the main perturbation of their orbital planes is due to the Earth's deviations from spherical symmetry and by far the main error in the measurement of frame-dragging using their orbits is due to the uncertainties in the Earth's even zonal spherical harmonics \cite{kau}. The Earth's gravitational field and its gravitational potential can be expanded in spherical harmonics and the even zonal harmonics are those harmonics of even degree and zero order. These spherical harmonics, denoted as $J_{2n}$, where ${2n}$ is their degree, are those deviations from spherical symmetry of the Earth's gravitational potential that are axially symmetric and that are also symmetric with respect to the Earth's equatorial plane: they produce large secular drifts of the nodes of the LAGEOS satellites. In particular, the flattening of the Earth's gravitational potential, corresponding to the second degree zonal harmonic $J_2$ describing the Earth's quadrupole moment, is by far the largest error source in the measurement of frame-dragging since it produces the largest secular perturbation of the node of LAGEOS \cite{ciu86,ciu96}. But thanks to the observations of the geodetic satellites, the Earth's shape and its gravitational field are extremely well known. For example, the flattening of the Earth's gravitational potential is today measured \cite{rolf} with an uncertainty of only about one part in $10^7$ that is, however, still not enough to test frame-dragging. To eliminate the orbital uncertainties due to the errors in the Earth's gravity models, the use of both LAGEOS and LAGEOS2 was proposed \cite{ciu96}. However, it was not easy to confidently assess the accuracy of some earlier measurements \cite{sci} of the Lense-Thirring effect with the LAGEOS satellites, given the limiting factor of the uncertainty of the gravity models available in 1998.

The problem of the uncertainties in the Earth's gravity field was overcome in March 2002 when the twin GRACE (Gravity Recovery And Climate Experiment) \cite{gra1,gra2} spacecraft of NASA were launched in a polar orbit at an altitude of approximately 400 km and about 200-250 km apart. The spacecraft range to each other using radar and they are tracked by the Global Positioning System (GPS) satellites. The GRACE satellites have greatly improved our knowledge of the Earth's gravitational field. Indeed, by using the two LAGEOS satellites and the GRACE Earth gravity models, the orbital uncertainties due to the modeling errors in the non-spherical Earth's gravitational field are only a few per cent of the Lense-Thirring effect \cite{ciu04,ciu10,ciu11}. The method to measure the Lense-Thirring effect is to use two {\it observables}, provided by the two nodes of the two LAGEOS satellites, for the two unknowns: Lense-Thirring effect and uncertainty in the Earth quadrupole moment $\delta J_2$ \cite{ciu96}.

In 2004, nearly eleven years of laser-ranging data were analyzed. This analysis resulted in a measurement of the Lense-Thirring effect with an accuracy \cite{ciu04,ciu07,ciu10,ciu11} of approximately 10\%. The uncertainty in the largest Earth's even zonal harmonic, that is the quadrupole moment $J_2$, was eliminated by the use of the two LAGEOS satellites. However, the main remaining error source was due to the uncertainty in the Earth even zonal harmonics of degree strictly higher than two and especially to the even zonal harmonic of degree four, i.e., $J_4$.

After 2004, other accurate Earth gravity models have been published using longer GRACE observations. The LAGEOS analyses have then been repeated with new models, over a longer period and by using different orbital programs independently developed by NASA Goddard, the University of Texas at Austin \cite{rie09} and the German GeoForschungsZentrum (GFZ) Potsdam. The most recent frame-dragging measurements \cite{ciu10,ciu11} by a team from the universities of Salento, Rome, Maryland, NASA Goddard, the University of Texas at Austin and the GFZ Potsdam, have confirmed the 2004 LAGEOS determination of the Lense-Thirring effect. No deviations from the predictions of General Relativity have been observed.

\section{The LARES Space Experiment}

In the test of frame-dragging using LAGEOS and LAGEOS 2, the main error source is due to the even zonal harmonic of degree four, $J_4$; such an error can be as large as 10\% of the Lense-Thirring effect \cite{ciu10b}. Thus, to much increase the accuracy of the measurement of frame-dragging, one would need to eliminate that uncertainty by using a further observable, i.e., by using a laser-ranged satellite additional to LAGEOS and LAGEOS 2.

LARES (LAser RElativity Satellite) is a laser-ranged satellite of the Italian Space Agency (ASI), see Fig. 4. It was launched successfully on the 13th February 2012 with the qualification flight of VEGA, the new launch vehicle of the European Space Agency (ESA), which was developed by ELV (Avio-ASI) \cite{ciu12,ciupao}. LARES, together with the LAGEOS and LAGEOS 2 satellites and the GRACE mission \cite{gra1,gra2}, will provide an accurate test of Earth's frame-dragging with uncertainty of about 1\% and other tests of fundamental physics \cite{ciu10b,ciu11,ciu13}. The Lense-Thirring drag of the orbital planes of the LARES is approximately 118 milliarcseconds per year corresponding, at the LARES altitude, to approximately 4.5 m/yr.

LARES has the highest mean density of any known object orbiting in the Solar System. It is spherical and covered with 92 retro-reflectors, and it has a radius of 18.2 cm. It is made of a tungsten alloy, with a total mass of 386.8 kg, resulting in a ratio of cross-sectional area to mass that is about 2.6 times smaller than that of the two LAGEOS satellites \cite{ciupao}. Before LARES, the LAGEOS satellites had the smallest ratio of cross-sectional area to mass of any artificial satellite, such a ratio is critical to reduce the size of the non-gravitational perturbations.

The LARES orbital elements are as follows: the semi-major axis is 7820 km, orbital eccentricity 0.0007, and orbital inclination 69.5$^{o}$. It is currently well observed by the global ILRS station network. The extremely small cross-sectional area to mass ratio of LARES, i.e. 0.00027, which is smaller than that of any other artificial satellite, and its special structure, a solid sphere with high thermal conductivity, ensure that the unmodeled non-gravitational orbital perturbations are smaller than for any other satellite, in spite of its lower altitude compared to LAGEOS. This behavior has been confirmed experimentally using the first few months of laser ranging observations \cite{ciu12}.

\begin{figure}
\centering \resizebox{0.45\textwidth}{!}{
  \includegraphics{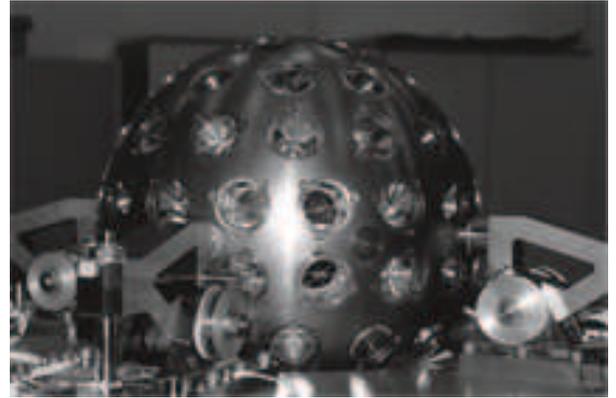}
} \caption{The LARES satellite (courtesy of the Italian Space Agency)}
\end{figure}

A number of papers have been published that analyze all the error sources, of both gravitational and non-gravitational origin, that can affect the LAGEOS and LARES experiments (see, e.g., \cite{ciu89,nasasi,riedis,ciu96,pet,ciupavper,ciu10,ciu10b,ciu11,ciu13}. The largest errors due to the uncertainties in the first two even zonal harmonics, of degree 2 and 4, i.e., $J_2$ and $J_4$, are eliminated using three observables, i.e. the three nodes of the LARES, LAGEOS and LAGEOS 2 satellites.
The error in the LARES experiment due to each even zonal harmonic up to degree 70 was analyzed in detail in \cite{ciu10b,ciu11}, the result is that the error due to each even zonal harmonic of degree higher than 4 is considerably less than 1\% of the Lense-Thirring effect and in particular the error is negligible for the even zonal harmonics of degree higher than 26. The LARES error analyses have been recently confirmed by a number of Monte Carlo simulations \cite{ciu13}.

We finally mention that in 2008, Smith et al. \cite{string} showed that String Theories of the type of Chern-Simons gravity, i.e., with action containing invariants built with the Riemann tensor squared, predict an additional drift of the nodes of a satellite orbiting a spinning body. Then, using the frame-dragging measurement obtained with the LAGEOS satellites, Smith et al. set limits on such String Theories that may be related to dark energy and quintessence. In particular, they have set a lower limit on the Chern-Simons mass that is related to more fundamental parameters, such as the time variation of a scalar field entering the Chern-Simons action, possibly related to a quintessence field. LARES will improve this limit on such fundamental physics theories. In summary, LARES will not only accurately test and measure frame-dragging but it will also provide other tests of fundamental physics.

\section{Conclusions}

We discussed time travels and clock `puzzles', mathematically predicted by Einstein's theory of General Relativity, and their experimental tests. Time travels should be divided in time travels to the future and time travels to the past. Time travel to the future, mathematically predicted by relativity, is indeed possible. It is just a consequence of the so-called twin `paradox', or twin `puzzle', of Special Relativity and of the time dilation in a gravitational field predicted by General Relativity. Twin `paradox' and time dilation have numerous, very accurate, experimental tests, including the direct measurement of time dilation by NASA's Gravity Probe A with an accuracy of about 0.02 $\%$. Furthermore, whenever we use a satellite navigator we put to test the clock effects predicted by Special and General Relativity. Without the inclusion of the relativistic clock effects in the positioning of the navigation satellites, the error in our own position would amount to kilometers. We then briefly discussed the possibility of time travel to the past, that is suggested by some mathematical solutions of the field equations of General Relativity, and some of the problems and paradoxes associated with it. We finally described the phenomenon of frame-dragging predicted by General Relativity to occur in the vicinities of a rotating body and of a mass-energy current. Frame-dragging implies a number of fascinating phenomena on the flow of time and on clocks. Frame-dragging phenomena on clocks and electromagnetic waves are extremely small in the solar system and thus very difficult to be detected, however, frame-dragging around the rotating Earth can be experimentally tested on test-gyroscopes and test-particles. We then presented the results of the LAGEOS and LAGEOS 2 experiment that measured frame-dragging with an uncertainty of about 10\% and NASA's Gravity Probe B that tested it with an uncertainty of about 20 $\%$. Finally, we described the space experiment using the LARES satellite, successfully launched in February 2012 for an accurate test of frame-dragging at the 1 $\%$ level.

\section{Acknowledgement}

This work was performed under the ASI contract n. I/034/12/0.

\end{document}